\documentclass[preprint]{aastex}

\begin{document}

\title{Multiband optical polarimetry of the BL Lac object PKS~2155--304.
Intranight and long term variability.}

\author{L. Tommasi\altaffilmark{1}}
\affil{Dipartimento di Fisica, Universit\`a di Milano, I-20133
Milan, Italy}
\author{R. D\'{\i}az} \affil{Observatorio Astron\'omico de C\'ordoba, Universidad Nacional de C\'ordoba, AR-5000 C\'ordoba, Argentina}
\author{E. Palazzi\altaffilmark{1}, E. Pian} \affil {ITESRE-CNR, I-40129 Bologna, Italy}
\author{E. Poretti} \affil{Osservatorio Astronomico di Brera, I-23807 Merate, Italy}
\author{F. Scaltriti\altaffilmark{1}} \affil{Osservatorio Astronomico di Torino, I-10025 Pino Torinese, Italy}
\and
\author{A. Treves} \affil{Dipartimento di Scienze, Universit\`a dell'Insubria, I-22100 Como, Italy}

\bigskip

\altaffiltext{1}{Visiting Astronomer, Complejo Astron\'omico El
Leoncito operated under agreement between the Consejo Nacional de
Investigaciones Cient\'{\i}ficas y T\'ecnicas de la Rep\'ublica
Argentina and the National Universities of La Plata, C\'ordoba
and San Juan.}

\begin{abstract}
The polarized and total flux of the BL Lac object PKS~2155--304
were monitored intensively and simultaneously in the optical
UBVRI bands with the Turin photopolarimeter at the CASLEO 2.15~m
telescope during 4 campaigns in June, August, November 1998 and
August 1999. The effective observation time amounted to $\sim$47
hours. PKS~2155--304 showed a linear polarization percentage (P)
usually ranging between 3\% and 7\% and a polarization position
angle (PA) mainly between 70$^\circ$ and 120$^\circ$. The highest
temporal resolution of our observations, 15 minutes, is
unprecedented for polarimetric monitoring of this source, and has
allowed us to detect amplitude variations of the linear
polarization percentage from 6 to 7.5\% in time scales of hours.
In some nights the polarization percentage seems to increase
toward shorter wavelengths, however the polarized spectrum does
not vary significantly with time.  The most remarkable
variability event occurred on 1998 June 18, when the degree of
linear polarization decreased by more than a factor 2 in one day
in all bands, while the PA rotated by 90 deg.  This is consistent
with the presence of two emission components, of different
polarization degree and position angle. Intranight variability of
P and PA can be interpreted with small amplitude physical or
geometrical changes within the jet. Measurements of the circular
polarization over time intervals of days set upper limits of
0.2\%. Simultaneous photometry taken with the Turin
Photopolarimeter and with a CCD camera at C\'ordoba Astronomical
Observatory did not show light variations correlated with those
of the linearly polarized flux.
\end{abstract}


\section{Introduction}

Relativistic jets are thought to characterize the geometry of the
most powerful extragalactic sources, blazars and gamma ray bursts
(GRB). In both classes of sources, synchrotron radiation is
responsible for the spectrum production over up to ten decades of
energy, from radio to soft gamma-rays. Linear polarization, seen
in blazars at a level of 1\% to 30\%
\citep{Angel,Takalo91,Takalo92,Visvanathan98} and recently
detected also in the optical afterglow of two GRBs
\citep{Covino99,Wijers99,Rol00}, is the most compelling
indication of synchrotron radiation. Relativistic beaming
enhances the polarized non thermal jet radiation with respect to
the unpolarized emission of the host galaxy, resulting in a high
net polarization percentage. Therefore, short time scale
variations of the polarized light trace the dynamics of the
plasma inside the jet, and their monitoring yields the most
direct insight into the jet physics and powering source.

Studies of linear polarization of blazars have been conducted from
radio to UV wavelengths \citep{Saikia88, Gabuzda96,Allen} and
optical polarization level has been also adopted as a
characteristic feature to select blazar sources
\citep{Borra84,Impey88,Jannuzi,Fleming93,Jannuzi93}.  However,
variability of polarization of blazars is poorly investigated,
and rapid and intensive monitoring campaigns, even of the
brightest and mostly polarized sources, are largely lacking,
except for S5~0716+714 \citep{Impey00}.

The BL Lac PKS~2155--304 is a nearby blazar (z=0.116, see Falomo,
Pesce \& Treves 1993), one of the brightest at optical to X-ray
frequencies and its polarized optical light has been observed
over many years by \citet{Brindle}, \citet{Mead}, \citet{Smith91},
\citet{Smith92}, \citet{Allen}, \citet{Jannuzi},
\citet{Courvoisier} and \citet{Pesce}. The main results are a
polarization fraction P varying from 2 to 10\%, a position angle
PA preferentially oriented at about 100-120$^\circ$ and a
wavelength dependence of polarization (WDP) tending to be null or
slightly increasing toward the blue. Observations of UV
polarization with HST by \citet{Allen} yielded P=3\% and
PA=101$^\circ$, and confirmed a positive WDP (i.e. P decreasing
toward longer wavelengths) through the whole optical-UV range.
Since in PKS~2155--304 the synchrotron mechanism is responsible
for the spectrum up to hard X-rays, producing a spectral maximum
at UV/soft X-rays (``blue blazars", see Padovani \& Giommi 1995,
Fossati et al. 1998), correlation is expected of flux variations
from optical to X-ray wavelengths. This was observed in past
multiwavelength campaigns (Edelson et al. 1995; Urry et al.
1997). In particular, in May 1994 the polarized optical light
variations were faster than those of the total light, and better
correlated with the UV variations. This indicates that optical
polarization is an efficient tracer of synchrotron emission, and
its variations probe, in the most detailed way, the energy
production processes.

To explore variability of the polarization with time scales from
months down to less than an hour, we conducted 4 campaigns of
polarimetric monitoring of PKS~2155--304 during 1998 and 1999 at
the 2.15~m telescope of the Complejo Astronomico El Leoncito
(CASLEO, Argentina).  Linear and circular polarization data along
with photometric data have been collected simultaneously in
\mbox{UBVRI} bands with a maximum temporal resolution of 15
minutes.

A description of the observations and data reduction process is
given in \S~2, while \S~3 is devoted to the presentation of the
photometric and polarimetric results.  In \S~4 we compare our
findings with the historical polarization behavior of
PKS~2155--304, and discuss them in \S~5. Preliminary and partial
results were presented in \citet{Treves99}.

\section{Observations}

\subsection{Polarimetry}

The polarimetric data in UBVRI bands were gathered during 4
campaigns conducted at the CASLEO 2.15~m telescope equipped with
the Photopolarimeter of the Turin Observatory, for a total of
about 47 hours on source exposure time. A log of all the
observations is shown in Table~\ref{linpoltab}.

The instrument is a double channel chopping photopolarimeter
\citep{Scaltriti89}, based on the design sketched by
\citet{Piirola}. A half-wave or quarter-wave retarder plate (for
linear or circular polarimetry, respectively) is inserted in the
light beam coming from the telescope. The plate is rotated
through 8 positions by 22.5$^\circ$ steps to explore polarization
of incoming light. A subsequent calcite slab splits the light into
ordinary and extraordinary rays, with orthogonal polarization
planes. These two rays are alternatively selected by a rotating
chopper; afterwards they pass through a pair of identical
diaphragms. In passing through the calcite slab the
extraordinary/ordinary component from the sky is refracted on the
ordinary/extraordinary component from the (object + sky). Thus
both components from the sky pass both diaphragms and
polarization of the sky is eliminated. Finally, the light is
split by four dichroic plus bandpass filter combinations and sent
to five photomultiplier tubes. The response of each of the five
channels corresponds to one of the \mbox{UBVRI} bands. In this
way, truly multicolor simultaneous photometry and polarimetry is
obtained. It can be noticed that the image is focused on the
diaphragm wheel. So the light beam incident on the retarder plate
is converging at an angle of about 3$^\circ$, taking into account
the focal ratio of the telescope. The two light beams (ordinary
and extraordinary) become then parallel by means of a couple of
lenses before reaching the dichroic filters. An integration time
of 10 seconds for each retarder plate position was chosen for all
observations, giving a final maximum time resolution of about
3.5~minutes. However, in order to get an acceptable {\em signal
to noise} ratio, the measurements have been binned with 4 or 8
plate cycles bins (approximately 15 or 30 minutes) according to
the different quality of data in the various nights. For all the
observations a 15$^{\prime \prime}$.8 diameter diaphragm was
used. The determination of the instrumental polarization and
orientation of zero point of the retarder with respect to North
is performed with the repeated observation of null and high
polarization standard stars, respectively. For each campaign, the
adopted instrumental parameters are the average values obtained
from observations of the standard stars during the various nights
of that observing run. The instrumental polarization ranges
typically between 0.16\% and 0.36\% in the U band, and between
0.02\% and 0.09\% in the other bands. The position angle
corresponding to the zero point is typically
79.8$^\circ$-82.6$^\circ$ with respect to the North.

Data analysis is performed by dedicated routines. They allow
calculation of polarization and position angle from the eight
integrations made in different orientations of the retarder
plate. Error estimates are obtained both from photon statistics
and least-squares fit to the eight integrations. The sky
background level, measured at the beginning of each set of 4
cycles of full rotations of the retarder (each cycle being
constituted by eight steps of 22.5$^\circ$), is taken into account
interpolating between sky measurements if the observed object (or
the diaphragm) has not been changed, otherwise the first
background value is used. The averaging procedure uses weights
determined from the inverse square of the estimated error for each
observation. The error estimate is taken either from the
least-squares fit of the double cosine curves to the eight
integrations in the different positions of the retarder or from
the photon statistics, whichever is greater. This prevents
weights from having values larger than the theoretical maximum
corresponding to photon noise limit, which could happen if only
the least-squares fits with only eight data points in each
observation were relied upon. The uncertainties affecting our
average measures of instrumental polarization range typically
between 0.02\% and 0.1\% (the largest values being recorded in U
and I bands). These systematic uncertainties affect the
measurements of the polarization percentage of the observed
source by 0.07\% in the u band, 0.04\% in B, 0.02\% in V, 0.03\%
in R, and 0.05\% (I). The systematic error on the orientation of
the zero position of the retarder, is estimated to be about
0.5$^\circ$, which leads to an uncertainty of about 0.01\% in the
polarization percentage and about 1$^\circ$ in the position angle
of PKS~2155-304. We also checked the repeatability of the
measurements of P and PA of standard stars from night to night.
The scatter on PA values is less then 1$^\circ$, while the
scatter in P is of the order of 0.02\% in all bands. Systematic
errors are much smaller than statistical errors, and therefore
have been ignored in the following. Our reduction routines also
account for the depolarization due to the instrumental setup
(telescope + polarimeter). To estimate the efficiency of 100\%
polarized light detection, we used all the observations of highly
polarized standards carried out at CASLEO from 1994 to 1997. An
average correcting factor 1.03$\pm$0.04 was found to be necessary
to reproduce the catalogued values in all bands from the measured
ones and has been introduced in the reduction routines.

In Table~\ref{linpoltab} and in Fig.~\ref{plottot} we report P and
PA averaged over each night, for all campaigns. Uncertainties on
the data points given in brackets are the maximum between the
statistical error associated with the individual points and the
standard deviation within the night, in the specified photometric
band. Only for the night of 4 August 1999, due to the poor
weather conditions resulting in low \textit{signal to noise}
ratio, all the collected data have been averaged without binning
over the plate cycles, resulting in the weighted means and
statistical errors that are reported in Table~\ref{linpoltab}.
Similar reduction has been performed also on circular polarimetry
data. The resulting nightly means are reported in
Table~\ref{linpoltab} along with statistical uncertainties.

\subsection{Photometry}
Photometry from CASLEO observations can be derived, from the same
polarimetric data, by adding counts from ordinary and
extraordinary rays for each plate position to obtain the total
flux from the source, measured in the five \mbox{UBVRI} bands. In
order to correct the source flux for sky transparency or
atmospheric extinction variations during the integrations, we
observed reference stars 2 and 3 of \citet{Smith} many times per
night during the runs of June and August 1998 and August 1999.
However, photometry is much more sensitive than polarimetry to the
sky conditions, therefore meaningful magnitude measurements have
been obtained only for the clearest nights or intervals.
Table~\ref{fottab} shows a summary of the results.

To partially overcome the limitation inherent to the CASLEO
photometry, in 25-28 August 1998 and 1-5 August 1999 we performed
additional differential photometry in \mbox{BVRI} bands at the
1.54~m telescope located at Bosque Alegre Astrophysical Station
of the C\'ordoba Astronomical Observatory (Argentina) equipped
with a Photometrics 1024$\times$1025 pixels CCD camera
(Multifunctional Integral Field Spectrograph, D\'{\i}az et al.
1997, D\'{\i}az et al. 1999). The magnitudes of PKS~2155-304 and
of the standard stars labelled as 2 and 3 by \citet{Smith} were
determined by means of aperture photometry routines in the MIDAS
package. A circle with a diameter of 11$^{\prime \prime}$.4 was
used to measure BL Lac flux. The magnitude differences measured
between the two standard stars are very close to the standard
ones; this supports the similarity between the instrumental and
the standard systems. Hence, standard magnitudes of PKS~2155--304
were calculated by assuming B=12.73, V=12.04, R=11.64 and I=11.28
for the standard star 2, as reported by \citet{Smith}. Nightly
means are shown as a summary in Table~\ref{fottab}.

\section{Results}

\subsection{Linear polarization}
\begin{figure}
\epsscale{1.0} \plotone{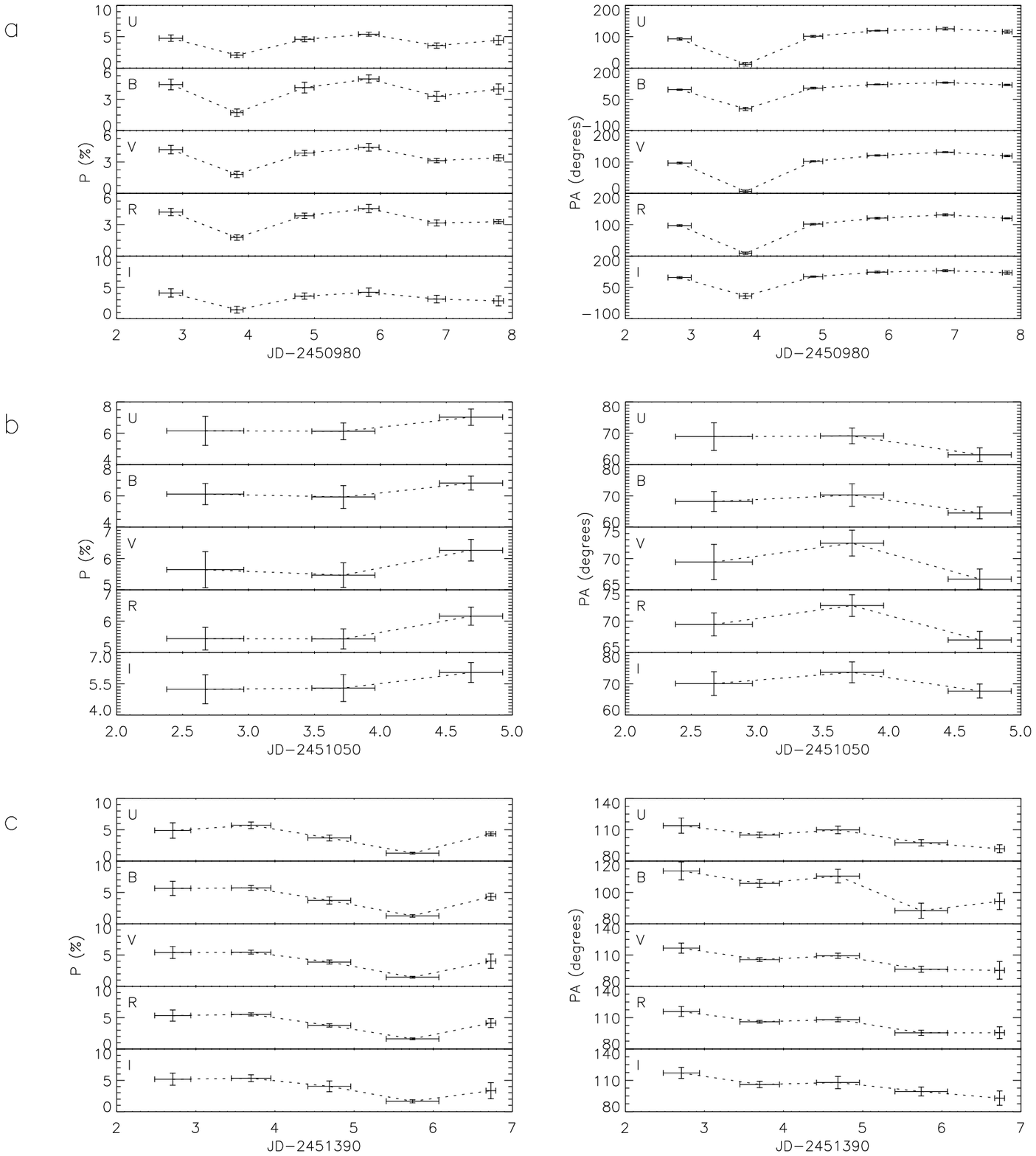} \caption{Nightly mean values
of P (left panels) and PA (right panels) in UBVRI band during
17-22 June 1998 (a), 26-28 August 1998 (b) and 1-5 August 1999
(c) campaigns. Horizontal bars represent the duration of the
observing run during each night. The dashed curves connecting the
points are meant only to guide the eye. \label{plottot}}
\end{figure}

\begin{figure}
\epsscale{0.51} \plotone{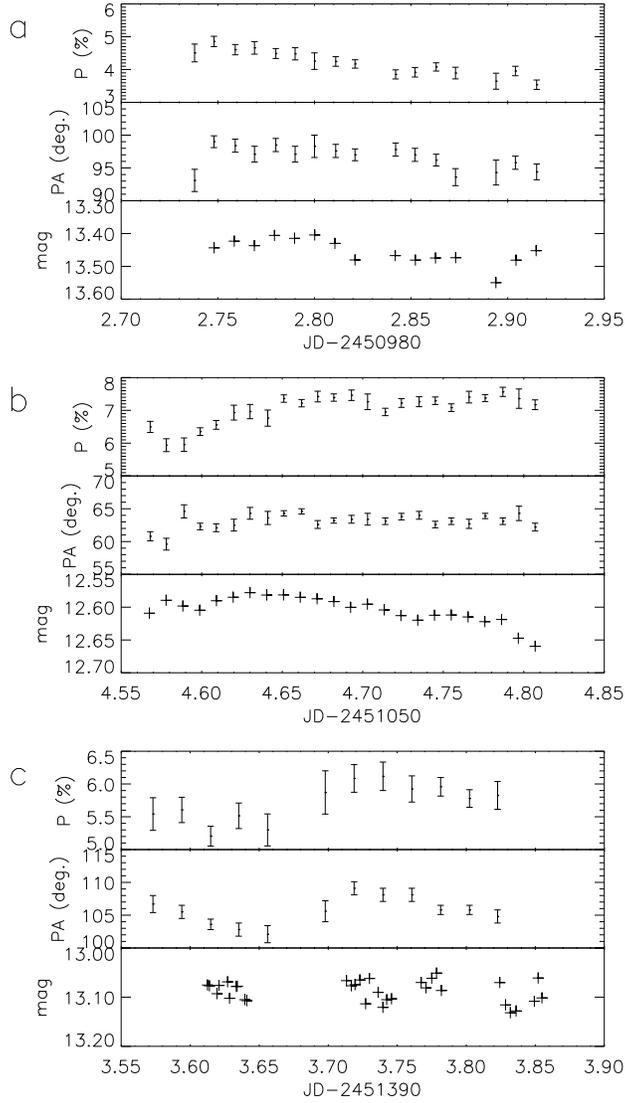} \caption{P, PA and
magnitude during 17 June 1998 (15 minutes bins in V band, panel
a), 28 August 1998 (15 minutes bins in U band, panel b) and 2
August 1999 (30 minutes bins in B band, panel c). Magnitudes are
from CASLEO (a, b) and C\'ordoba (c) observations. Observational
errors on photometric data can be deduced from Table~\ref{fottab}
\label{plotintranight}}
\end{figure}

The linear polarization percentage was roughly at the same average
level in all bands in June 1998 and August 1999, while it was
$\sim$1.3-1.5 times higher in August 1998 (see
Table~\ref{linpoltab}). Correspondingly, the polarization angle
was around 100-120$^\circ$ in June 1998 and August 1999, and at
70$^\circ$ in August 1998 (Fig.~\ref{plottot}). Day-to-day
variations generally did not exceed a factor $\sim$1.2 in
amplitude for P and $\sim$20$^\circ$ in PA, with the exception of
a relevant variability episode on 1998 June 18: the polarization
decreased in all bands (the largest variation was from 4 to 1.5\%
in I band) in no more than 21 hours, and subsequently returned in
22 hours to the value it had before the ``dip"
(Fig.~\ref{plottot}, panel a). Although a precise determination
of the halving and doubling time scales of this variation is
hampered by the day-time gap, based on the modest decreasing
trend present in all bands during the night of June 17 (see
Fig.~\ref{plotintranight} panel a) we estimate that the
halving/doubling time scale could be as short as 14 hours.
Simultaneously, the polarization plane rotated by about
90$^\circ$. Following our standard procedure, during June 1998
run we observed several high polarization standard stars, some of
them were monitored each night. Their position angles show
constant values in all the nights with a maximum deviation of
$\sim$1$^{\circ}$. We are therefore confident that the behaviour
of PKS~2155--304 cannot be attributed to a systematic
instrumental effect. A hint of a similar variability episode is
contained in August 1999 data, when the polarization decreased
from 5.5 to 1.5 \% in V band over two days, while no simultaneous
rotation of the PA was recorded. However, due to the poor quality
of the sky in these nights, this event should be considered with
some caution.

Intranight variations of up to a factor $\sim$1.3 in P and
$\sim$7$^{\circ}$ in PA are sometimes detected  with time scales
of few hours. These are either random, or they follow the longer
time scale trends (as in the night of 17 June 1998, where the
long time trend is prevailing). Data from the three nights in
which such variations are more apparent (one for each campaign)
are plotted for selected bands in Fig.~\ref{plotintranight}. In
the same nights, similar behaviors are recorded also in the other
bands. The event recorded during 28 August 1998 (see
Fig.~\ref{plotintranight} panel b) is particularly notable for
smoothness of the temporal profile. After a small decrease in
polarization in the first 15 minutes to a low state, maintained
for about half an hour, in the next 2 hours the source exhibits a
regular and rather smooth increase up to 1.2 times the minimum
value. During other nights, the variations detected over similar
time scales are more modest or absent.

\begin{figure}
\epsscale{1.0} \plotone{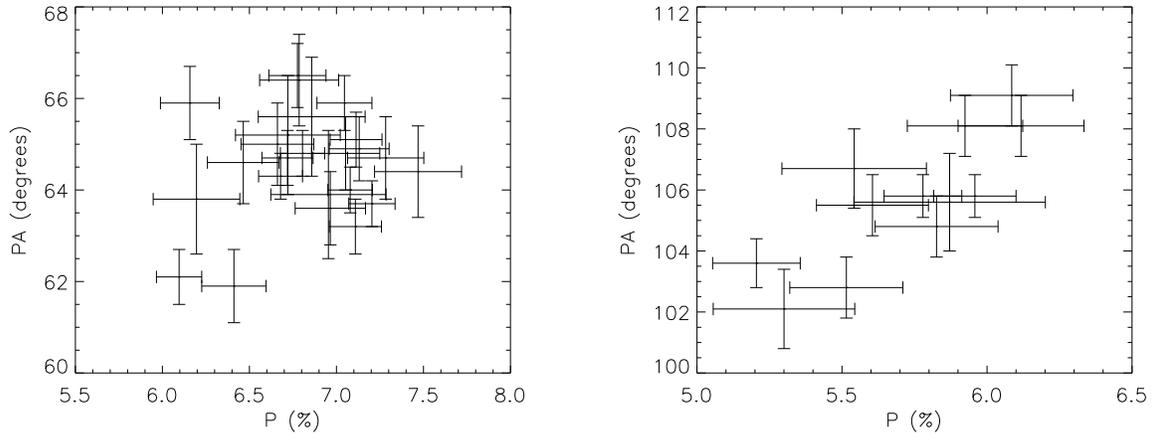} \caption{PA vs P in B band
during 28 August 1998 with 15 minutes bins (left panel) and
during 2 August 1999 with 30 minutes bins (right panel).
\label{ppatot}}
\end{figure}

We searched for a correlation between polarization and position
angle on various time scales. In some nights there is evidence of
such a correlation, while in other cases the variations in P and
PA seem little correlated (Fig.~\ref{ppatot}). Moreover, within
the same night, the P-PA correlation may be different in the five
photometric bands. Looking at inter-night variations, a
correlation can be evidenced only for the dip of 18 June 98
(Fig.~\ref{plottot} panel a). Finally, looking at the whole data
set (see Table~\ref{linpoltab}), there is suggestion of a
correlation between subsequent campaigns, in the sense of
increasing P and decreasing PA from June to August 1998 and
viceversa from August 1998 to August 1999.

\subsection{Wavelength dependence of polarization}

\begin{figure}
\epsscale{0.4} \plotone{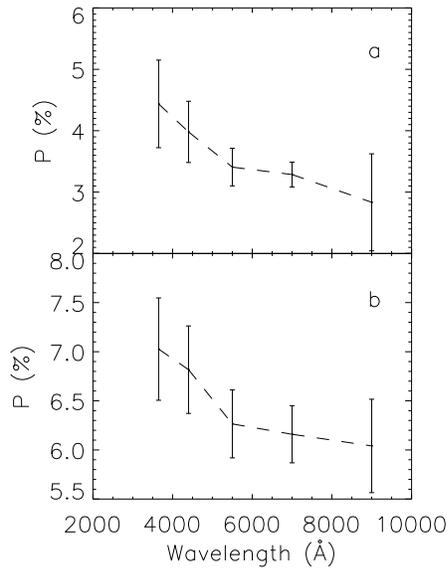} \caption{Wavelength
dependence of polarization for the nights of 22 June 1998 (a) and
28 August 1998 (b). The dashed curves connecting the points are
meant only to guide the eye. \label{wdp}}
\end{figure}

We also investigated the wavelength dependence of polarization
(WDP). Examples from selected nights are in Fig.~\ref{wdp}.
Non-null WDP was found in some nights during the runs of June and
August 1998, in the sense of decreasing polarization with
increasing wavelength. This dependence disappeared in the run of
August 1999, when the polarized spectrum appears flat. Following
\citet{Smith91}, we defined the parameters
\begin{displaymath}
P_\nu=\frac{d\log P}{d\log \nu} \; ,
\end{displaymath}
and
\begin{displaymath}
{\it PA}_\nu=\frac{d {\it PA}}{d\log \nu} \; .
\end{displaymath}
which describe the spectral shape of the polarization percentage
and position angle, and determined them by fitting power-laws to
the daily averages of the UBVRI linear polarization percentages
and position angles. Table~\ref{dlogp} reports the values of
$P_\nu$ and ${\it PA}_\nu$, with their standard deviations as
resulting from the fit. While in some nights $P_\nu$ is obtained
with more than 2$\sigma$ confidence, the fitted ${\it PA}_\nu$
values are poorly determined. The suggestion is that the best
determined ones (27 and 28 August 1998) are negative, namely the
PA increases toward the red wavelengths.

\subsection{Circular polarization}

Results on circular polarization are reported in
Tab~\ref{linpoltab}. Generally, no significant detection is
present; marginally significant (slightly less than 3$\sigma$)
circular polarization is seen in \mbox{UBV} bands on 1998
November 19 at a level consistent with -0.2\% in all filters,
within 1$\sigma$.

No clear pattern is deduced from comparison of values in different
filters in the other nights, nor is it discerned any preferred
value, in a given filter, by comparison among different nights.
Conservatively, we set an upper limit of 0.2\% at the 3~$\sigma$
level for our measurement of circular polarization of
PKS~2155--304.

\subsection{Correlation between polarimetry and photometry}

\begin{figure}
\epsscale{0.6} \plotone{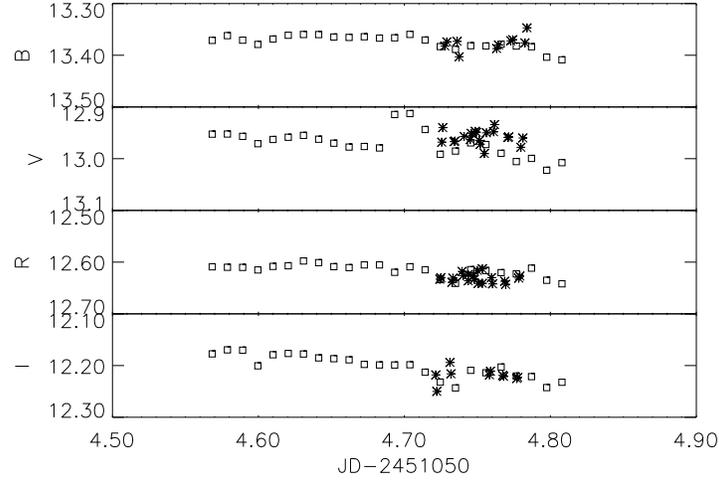} \caption{Magnitude
measurements for the night of 28 August 1998. Squares represent
CASLEO data, asterisks are CCD measurements. For the
uncertainties on the data points see Tab~\ref{fottab}.
\label{plotintranight1}}
\end{figure}

The CASLEO photometry agrees well with the C\'ordoba photometry,
as seen in Fig.~\ref{plotintranight1}. Variability between
successive runs (on months intervals) is always present and
amounts to some tenths of a magnitude, with no apparent
correlation with polarization. Variations within the same night
and from night to night in the same run are observed at a level up
to $\sim$10\%, and will be discussed in E. Poretti et al. (in
preparation).

Generally, variations of the total and polarized flux are not
found to be correlated within the same run. In particular, the
sharp decrease of the linear polarization which occurred on 18
June 1998 had no counterpart in the CASLEO photometry (see
Tab~\ref{fottab}), and the significant decrease of polarization
between 1999 August 2 and August 3 corresponds to practically
unaltered photometric state, as shown by the CASLEO and C\'ordoba
photometry in Table~\ref{fottab}. Similarly, no correlation is
found on intranight scales. In particular, during the night of
1998 August 28, no correlated variation in total flux has been
recorded simultaneous with the strong variability of polarization
percentage in the early portion of the run (see
Fig.~\ref{plotintranight}, panel b). The smooth and broad maximum
present in the total light curve in the central part of the night
cannot be reliably correlated with the fast variation seen in our
polarimetric data.

\section{Comparison with previous polarimetric results}

\begin{figure}
\epsscale{0.6} \plotone{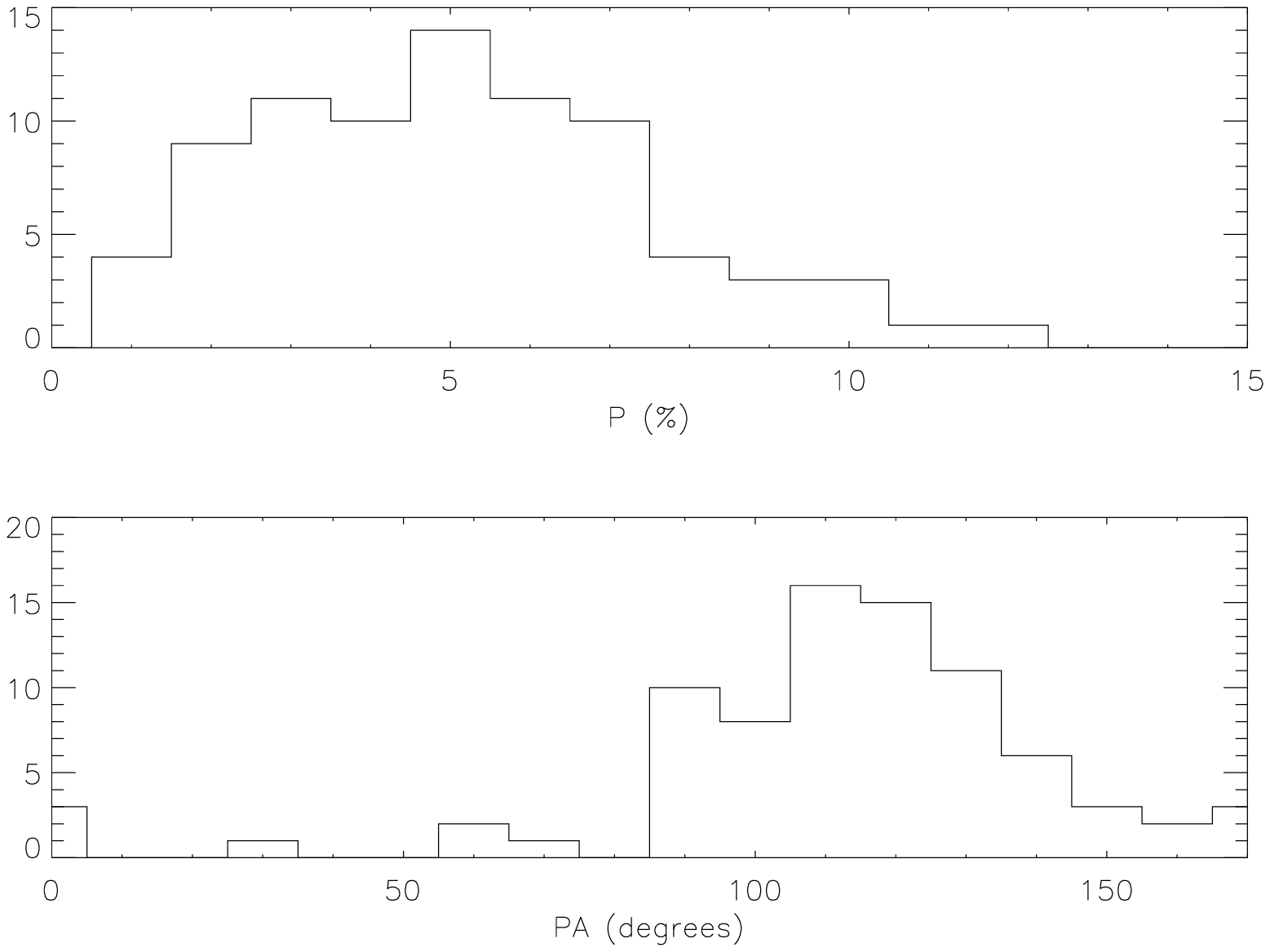} \caption{Histograms of P
and PA data of PKS 2155--304 in the V band collected from the
literature
\citep{Brindle,Mead,Smith91,Smith92,Allen,Jannuzi,Courvoisier,Pesce}.
\label{apjlett3}}
\end{figure}

As mentioned above, PKS~2155--304 has been the target of various
optical polarimetric campaigns in previous years. We compared our
results with those reported in the literature by representing in
histograms P and PA values from previous measurements and from
our own (Fig.~\ref{apjlett3}). Both distributions exhibit a broad
peak, indicating some historical preferred values. The peak of
polarization percentage occurs at about 5.5\%. The distribution
is strongly asymmetric, with an apparent tail toward values
larger than 10\%. PA values are mostly included in the interval
90$^{\circ}$-160$^{\circ}$, with a mode $\sim$115$^{\circ}$.
However many strongly deviating values were recorded, both in the
past and in our observations. In particular all the PAs lying
between 50$^{\circ}$ and 80$^{\circ}$, as well as one of those
smaller than 10$^{\circ}$, were observed during our campaigns.

\section{Discussion}

We have obtained polarimetric data of PKS~2155--304 over long time
intervals at an unprecedentedly high sampling rate. This allowed
us to study in detail the long and short term polarization
variability of the source, in particular the intranight events,
and the shape of the polarized spectrum.

Possible dilution of the polarization due to the host galaxy of
PKS 2155--304, a giant elliptical, is expected to be wavelength
dependent, because its spectrum should be significantly redder
than that of the nucleus.  In fact, given the measured magnitudes
of the host galaxy ($M_H=-26.8$, Kotilainen Falomo \& Scarpa 1998
and $M_I=-24.2$ Falomo et al. 1991), using the color indexes of a
typical elliptical galaxy reported by \citet{Fukugita95},
\citet{Kotilainen98} and \citet{Bertone00} we calculated the host
galaxy flux $F_{host}$ in the \mbox{UBVRI} bands. Given the
diameter of the projected aperture size for photopolarimetry
(15$^{\prime \prime}$.8) and the effective de Vaucouleurs radius
for the host galaxy reported in \citet{Falomo91} (4$^{\prime
\prime}$.5 in I Gunn filter) and \citet{Kotilainen98} (1$^{\prime
\prime}$.75 in H band), we assume that 100\% of the galaxy flux
$F_{host}$ is included in our photometry. Then, by subtracting it
from the total LIGHT (AGN+galaxy) represented by the magnitudes in
Table~\ref{fottab}, we computed the AGN flux $F_{AGN}$. Finally,
in the hypothesis that the host galaxy only dilutes the
polarization percentage of the AGN with its total flux
contribution, and ignoring depolarization by Faraday rotation, we
estimated the intrinsic polarization of the AGN light according to
\begin{displaymath}
P_{int} = P_{meas}  \left(1 + \frac{F_{host}}{F_{AGN}} \right)
\end{displaymath}
where $P_{meas}$ are from Table~\ref{linpoltab}. The result for
the night of 28 August 1998, one of the nights with best
determined WDP, is a constant polarization P=7.1\% over the
\mbox{UVBRI} bands. Observed values of $P_\nu$ smaller than that
of 28 August 1998 have all larger errors (see Table~\ref{dlogp}),
and therefore the corresponding corrected $P_\nu$ indices are not
significantly below zero (except the marginally significant case
of 3 August 1999). Therefore, no strong evidence is present of
negative WDP in our data. We note that our aperture includes also
light from a galaxy located at 4$^{\prime \prime}$ from the BL
Lac nucleus (G1, Falomo, Pesce \& Treves 1993). However its
emission (m$_{R} \sim 20$) does not contribute significantly to
the depolarization. We also neglected any possible polarization
induced by dust scattering in the host, this being usually
several orders of magnitude smaller than the AGN synchrotron
polarization.

Past studies \citep{Smith91} based on WDP have excluded that an
accretion disk is contributing significantly to the optical emission of
PKS~2155--304, in which case higher polarization would be expected at
longer wavelengths. Similar conclusions were reached by \citet{Urry93}
based on the correlated behavior of the optical and UV light curves.
Our finding of a flat polarized spectrum is consistent with the
conclusions of \citet{Smith91}.

As proposed to explain optical polarization of GRB afterglows
\citep{Gruzinov99,Medvedev99}, which exhibit polarization levels
comparable to this and other blazars, one can envisage two
scenarios: one with a highly ordered magnetic field, and one with
highly tangled field. The two cases can be distinguished basing
on position angle measurements: a constant PA indicating the
former case, a variable PA the latter. Our measurements of
PKS~2155--304, combined with the historical ones, suggest the
following scenario. A component endowed with a regular magnetic
field, usually dominant, is responsible for the preferred
PA$\sim$115$^{\circ}$.  The event recorded on 18 June 1998, with
a sudden decrease of P and rotation of the PA, suggests, instead
of a large and rapid rotation of the magnetic field axis, a
quenching of the ordered component, and the emerging of a
different component, with a more tangled field. The lower level
of polarization in this case can be explained with the presence
of $n$ sub-regions, or ``patches", each with size comparable to
the coherence length of the magnetic field, so that the net P is
given by P$_{max}$/$\sqrt{n}$, where P$_{max}$ is the
polarization of each individual patch
\citep{Gruzinov99,Wijers99}. From the measurement of P in low
state ($\sim$1\%) and from the expected polarization of the
synchrotron radiation ($\sim$70\%) one deduces a number of patches
of $\sim$5000. The ordered-field component could be associated
with a jet observed slightly off-axis, as envisaged by
\citet{Covino99,Ghisellini99} and \citet{Medvedev99}. However,
also the rise of a single component with orthogonal polarization
plane could explain the sudden decrease in P accompanied by a
90$^\circ$ rotation of the PA.

Smaller time scale (intranight) variations of P can be interpreted
with physical changes within the jet (modifications of the
electron energy distribution law due to a propagating shock or
small variations in magnetic field strength), while short term
variations in PA can be due to geometric variations (varying
orientation of jet or magnetic field vector). Note that a detailed
model should account for the absence of correlation between the
polarized and total flux that is shown in the observations
reported here.

Previous reports of optical circular polarization measurements in
blazars include marginal detections by \citet{Takalo93} for
3C~66A, OJ~287, and  by \citet{Valtaoja93} for PKS~0735+178 and
PKS~0422+004.  For PKS~2155--304, only measurements with HST in
the UV are reported \citep{Allen}. The absence of significant
circular polarization, in a simple and homogeneous scenario,
agrees with expectation that this should be a factor $\gamma_e$
less than the linear P, where $\gamma_e$ is the Lorentz factor of
the electrons, therefore out of reach of our accuracy, given that
in blazars typically $\gamma_e
> 10^3$. More complex scenarios, with inhomogeneous jets and
helical magnetic fields \citep{Valtaoja84} foresee larger relative
values of circular polarization with respect to linear
polarization. This could be invoked if the hint of circular
polarization detection of $\sim$0.2\% on 1998 November 19 turned
out to be real.

Since the polarized flux in PKS~2155--304 is produced by
synchrotron radiation in the nuclear region, with very small
dilution from the host galaxy, we expect P variations to be well
correlated with those of the X-ray flux (see also Smith et al.
1992). For the same reason, we do not expect radio polarization
variations to be correlated with the optical polarization,
because the former originate in more extended regions, external
to the inner nucleus. Simultaneous monitoring of PKS~2155--304 at
optical to X-ray frequencies could help discerning the exact
causes of small time scale variability and advance our
understanding of the physical behavior of ``blue", synchrotron
dominated, blazars.

\acknowledgments We thank G. Gimeno at Bosque Alegre
Astrophysical Station for collaboration. We are also grateful to
G. Ghisellini for a critical reading of the manuscript and to B.
Wills for a helpful and constructive referee report. Financial
support from EC grant ERBFMRXCT 98-0195 and Italian MURST COFIN
98021541 are acknowledged.

\clearpage

\clearpage

\begin{deluxetable}{ccccccc}
\tablecolumns{6} \tabletypesize{\scriptsize} \tablewidth{0pt}
\tablecaption{Journal of Observations \label{linpoltab}}
\tablehead{ \colhead{Date} & \colhead{JD \tablenotemark{a}} &
\colhead{On source integration} & \colhead{Pol. State
\tablenotemark{b}} & \colhead{Filter} & \colhead{P (\%)} &
\colhead{PA ($^{\circ}$)}
\\
\colhead{} & \colhead{} & \colhead{time (min.)} & \colhead{} &
\colhead{} & \colhead{} & \colhead{} }
\startdata Jun 1998& & &
&& \\*

17&2450982.726&150& l &U& 4.76 (0.50)& 93.1 (3.9)\\*
  &2450982.920&   &   &B& 4.43 (0.51)& 96.4 (2.9)\\*
  &&   &   &V& 4.19 (0.41)& 96.7 (2.7)\\*
  &&   &   &R& 4.21 (0.35)& 97.1 (2.7)\\*
\smallskip
  &&   &   &I& 4.10 (0.66)& 96.5 (4.6)\\

18&2450983.767&75 & l &U& 2.04 (0.38)& 12.0 (5.8)\\*
  &2450983.870&   &   &B& 1.72 (0.36)&  4.3 (7.4)\\*
  &&   &   &V& 1.81 (0.30)&  6.4 (3.8)\\*
  &&   &   &R& 1.76 (0.25)&  8.6 (3.7)\\*
\smallskip
  &&   &   &I& 1.42 (0.54)&  8.3 (11.3)\\

19&2450984.777&150& l &U& 4.58 (0.40)& 101.3 (3.0)\\*
  &2450984.922&   &   &B& 4.12 (0.51)& 104.1 (3.9)\\*
  &&   &   &V& 3.86 (0.25)& 102.4 (2.0)\\*
  &&   &   &R& 3.84 (0.25)& 101.5 (2.3)\\*
\smallskip
  &&   &   &I& 3.59 (0.50)& 101.0 (3.4)\\

20&2450985.751&120& l &U& 5.70 (1.29)& 118.8 (6.0)\\*
  &2450985.904&   &   &B& 5.06 (0.51)& 121.6 (3.1)\\*
  &&   &   &V& 4.37 (0.55)& 120.9 (2.9)\\*
  &&   &   &R& 4.46 (0.60)& 120.5 (3.4)\\*
\smallskip
  &&   &   &I& 4.04 (0.96)& 121.1 (5.9) \\

21&2450986.791&135& l &U& 3.58 (0.43)&  125.4 (4.0)\\*
  &2450986.926&   &   &B& 3.29 (0.47)&  130.1 (3.2)\\*
  &&   &   &V& 3.13 (0.21)&  131.5 (1.6)\\*
  &&   &   &R& 3.16 (0.29)&  131.3 (3.0)\\*
\smallskip
  &&   &   &I& 3.13 (0.60)&  129.6 (4.7) \\

22&2450987.760&75 & l &U& 4.44 (0.72)&  115.9 (4.6)\\*
  &2450987.833&   &   &B& 3.98 (0.50)&  119.1 (3.3)\\*
  &&   &   &V& 3.41 (0.31)&  119.4 (2.6)\\*
  &&   &   &R& 3.29 (0.20)&  120.1 (1.8)\\*
\smallskip
  &&   &   &I& 2.83 (0.79)&  120.0 (7.7) \\
Aug 1998& & & & & \\*

26&2451052.516&345& l &U& 6.15 (0.93)&  68.9 (4.4)\\*
  &2451052.816&   &   &B& 6.12 (0.68)&  68.2 (3.2)\\*
  &&   &   &V& 5.64 (0.58)&  69.4 (2.8)\\*
  &&   &   &R& 5.44 (0.36)&  69.5 (1.8)\\*
\smallskip
  &&  &    &I& 5.24 (0.69)&  70.1 (3.8)  \\

27&2451053.590&240& l &U& 6.12 (0.54)&  69.1 (2.5)\\*
  &2451053.836&   &   &B& 5.93 (0.73)&  70.3 (3.6)\\*
  &&   &   &V& 5.47 (0.39)&  72.4 (2.1)\\*
  &&   &   &R& 5.43 (0.32)&  72.5 (1.7)\\*
\smallskip
  &&   &   &I& 5.30 (0.64)&  73.7 (3.3)\\

28&2451054.562&330& l &U& 7.03 (0.52)&  63.1 (2.2)\\*
  &2451054.808&   &   &B& 6.82 (0.44)&  64.5 (1.9)\\*
  &&   &   &V& 6.26 (0.34)&  66.7 (1.6)\\*
  &&   &   &R& 6.16 (0.29)&  67.0 (1.3)\\*
\smallskip
  &&   &   &I& 6.04 (0.47)&  67.7 (2.5)\\
Nov 1998& & & & & \\*

15&2451133.527& 22& c &U& 0.19 (0.16) & - \\*
  &2451133.539&   &   &B& 0.12 (0.08) & - \\*
  &&   &   &V&-0.07 (0.05) & - \\*
  &&   &   &R&-0.02 (0.08) & - \\*
\smallskip
  &&   &   &I& 0.04 (0.09) & - \\

16&2451134.547& 15& c &U&-0.08 (0.10)& - \\*
  &2451134.555&   &   &B& 0.25 (0.11) & - \\*
  &&   &   &V&-0.10 (0.07) & - \\*
  &&   &   &R&-0.10 (0.05)& - \\*
\smallskip
  &&   &   &I&-0.05 (0.11) & - \\

18&2451136.547& 15& c &U&-0.04 (0.08)& - \\*
  &2451136.555&   &   &B&0.17 (0.11) & - \\*
  &&   &   &V&0.13 (0.05) & - \\*
  &&   &   &R&0.02 (0.05) & - \\*
\smallskip
  &&   &   &I&-0.14 (0.10) & - \\

19&2451137.551& 15& c &U&-0.20 (0.08)& - \\*
  &2451137.558&   &   &B& -0.32 (0.13)& - \\*
  &&   &   &V& -0.15 (0.05) &- \\*
  &&   &   &R& -0.07 (0.05) &- \\*
\smallskip
  &&   &   &I& -0.19 (0.10) & - \\
Aug 1999& & & & & \\*

 1&2451392.594&270& l &U& 4.88 (1.23)&  113.8 (7.3)\\*
  &2451392.828&   &   &B& 5.64 (1.13)&  113.7 (5.7)\\*
  &&   &   &V& 5.41 (0.95)&  116.6 (4.9)\\*
  &&   &   &R& 5.34 (0.89)&  115.9 (4.7)\\*
\smallskip
  &&   &   &I& 5.18 (0.97)&  117.1 (5.3)\\

 2&2451393.570&285& l &U& 5.70 (0.51)&  105.0 (2.6)\\*
  &2451393.820&   &   &B& 5.71 (0.39)&  105.7 (2.5)\\*
  &&   &   &V& 5.46 (0.31)&  105.6 (1.8)\\*
  &&   &   &R& 5.53 (0.23)&  106.1 (1.3)\\*
\smallskip
  &&   &   &I& 5.33 (0.55)&  106.0 (3.0)\\

 3&2451394.566&300& l &U& 3.68 (0.46)&  109.8 (3.5)\\*
  &2451394.820&   &   &B& 3.70 (0.57)&  110.4 (4.3)\\*
  &&   &   &V& 3.88 (0.33)&  109.3 (2.4)\\*
  &&   &   &R& 3.76 (0.26)&  108.1 (2.1)\\*
\smallskip
  &&   &   &I& 4.04 (0.86)&  107.8 (5.8)\\

 4&2451395.574&285& l &U& 1.26 (0.14)&   97.5 (3.1)\\*
  &2451395.918&   &   &B& 1.25 (0.21)&   88.3 (4.8)\\*
  &&   &   &V& 1.46 (0.14)&   96.4 (2.8)\\*
  &&   &   &R& 1.62 (0.14)&   95.5 (2.5)\\*
\smallskip
  &&   &   &I& 1.66 (0.25)&  99.3 (4.3) \\

 4&2451395.676& 60& c &U&-0.15 (0.06)& - \\*
  &2451395.715&   &   &B& 0.01 (0.10) & - \\*
  &&   &   &V&-0.15 (0.05) & - \\*
  &&   &   &R& 0.08 (0.03) & - \\*
\smallskip
  &&   &   &I&-0.15 (0.09) & - \\

 5&2451396.703& 75& l &U& 4.34 (0.32)&    91.8 (3.8)\\*
  &2451396.762&   &   &B& 4.31 (0.56)&    94.3 (5.3)\\*
  &&   &   &V& 4.03 (1.13)&    95.3 (8.5)\\*
  &&   &   &R& 4.14 (0.73)&    95.6 (5.6)\\*
\smallskip
  &&   &   &I& 3.35 (1.29)&  92.9  (6.9)   \\
\enddata
\tablenotetext{a}{JD at the first and last measurement of each
night.} \tablenotetext{b}{l=linear polarization; c=circular
polarization.}
\end{deluxetable}

\clearpage

\begin{deluxetable}{cccccccccccccc}
\tablecolumns{14} \tabletypesize{\scriptsize} \tablewidth{0pt}
\tablecaption{Photometry data (nightly means) \label{fottab}}
\tablehead{ \colhead{}    &  \colhead{m$_{U}$} & \colhead{} &
\multicolumn{2}{c}{m$_{B}$} & \colhead{}&
\multicolumn{2}{c}{m$_{V}$} & \colhead{}&
\multicolumn{2}{c}{m$_{R}$} & \colhead{}&
\multicolumn{2}{c}{m$_{I}$}\\
\cline{4-5} \cline{7-8} \cline{10-11} \cline{13-14 }\\
\colhead{Date} & \colhead{PMT}& \colhead{}& \colhead{PMT} &
\colhead{CCD}& \colhead{}& \colhead{PMT} & \colhead{CCD}
&\colhead{}  & \colhead{PMT}   & \colhead{CCD} &\colhead{}&
\colhead{PMT} & \colhead{CCD} } \startdata
17 Jun 1998&    13.03& & 13.87& & &13.45& & &13.09& & &12.65& \\
\smallskip
 &(0.02)&&(0.01)&&&(0.04)&&&(0.03)&&&(0.08)&\\
18 Jun 1998&    12.96& & 13.79& & &13.36& & &13.05& & &12.61& \\
\smallskip
 &(0.04)&&(0.04)&&&(0.04)&&&(0.04)&&&(0.04)&\\
19 Jun 1998&    12.86& & 13.68& & &13.28& & &12.92& & &12.47& \\
\smallskip
 &(0.02)&&(0.03)&&&(0.04)&&&(0.03)&&&(0.04)&\\
25 Aug 1998& & & &13.48& & &13.10& & & 12.78& & & 12.35 \\
\smallskip
 & & & &(0.09)&&&(0.09)&&&(0.08)&&&(0.09)\\
26 Aug 1998&    12.41& & 13.36&13.39& &13.07&12.99& &12.58&12.67& &12.15&12.25 \\
\smallskip
 &(0.06)&&(0.05)&(0.03)&&(0.06)&(0.01)&&(0.05)&(0.01)&&(0.05)&(0.02)\\
27 Aug 1998&    12.55 & & 13.32 &13.40& &12.88 &13.00& &12.55 &12.66& &12.10 &12.25 \\
\smallskip
 &(0.09)&&(0.09)&(0.03)&&(0.10)&(0.01)&&(0.09)&(0.01)&&(0.09)&(0.02)\\
28 Aug 1998&    12.60 & & 13.37 &13.38& & 12.97 &12.96& &12.61 &12.63& &12.20 & 12.22 \\
\smallskip
 &(0.02)&&(0.02)&(0.03)&&(0.03)&(0.01)&&(0.02)&(0.01)&&(0.03)&(0.02)\\
1 Aug 1999& 12.10 & &13.00 &12.98& &12.59 &12.56& &12.28 &12.24& &11.83 & 11.85\\
\smallskip
 &(0.13)&&(0.07)&(0.05)&&(0.04)&(0.03)&&(0.03)&(0.02)&&(0.03)&(0.02) \\
2 Aug 1999&       & &      &13.09& &      &12.67& &      &12.34& &      & 11.96 \\
\smallskip
 &      &&      &(0.02)&&      &(0.02)&&      &(0.01)&&      &(0.02)\\
3 Aug 1999&       & &      &13.09& &      &12.68& &      &12.35& &      & 11.97 \\
\smallskip
 &      &&      &(0.02)&&      &(0.01)&&      &(0.01)&&      &(0.02)\\
4 Aug 1999& & & &13.13& & &  12.72& & & 12.39& & &12.03 \\
\smallskip
 &&&&(0.02)&&&(0.08)&&&(0.05)&&&(0.02)\\
5 Aug 1999& & & &13.19& & & 12.77& & & 12.44& & &12.07 \\
 &&&&(0.05)&&&(0.02)&&&(0.04)&&&(0.03)\\
\enddata
\tablecomments{PMT denotes measurements performed with the
photopolarimeter from CASLEO, while CCD denotes measurements
performed with CCD from C\'ordoba Astronomical Observatory.
Numbers in brackets represent the standard deviation associated
with the nightly means, which in several cases is dominated by the
intranight variability of the source.}
\end{deluxetable}

\clearpage

\begin{deluxetable}{ccc}
\tablecolumns{3} \tabletypesize{\scriptsize} \tablewidth{0pt}
\tablecaption{Results from WDP analysis. \label{dlogp}}
\tablehead{ \colhead{Date}  &\colhead{$P_\nu$}  &\colhead{${\it
PA}_\nu$}   } \startdata
17 Jun 1998 & 0.16$\pm$0.16 & -3.19$\pm$5.22    \\
18 Jun 1998 & 0.22$\pm$0.30 &  1.80$\pm$9.18    \\
19 Jun 1998 & 0.25$\pm$0.14 &  1.19$\pm$4.35    \\
20 Jun 1998 & 0.29$\pm$0.13 & -3.00$\pm$3.70    \\
21 Jun 1998 & 0.15$\pm$0.18 & -4.33$\pm$5.54    \\
22 Jun 1998 & 0.42$\pm$0.20 & -4.08$\pm$5.59    \\
26 Aug 1998 & 0.20$\pm$0.17 & -1.72$\pm$4.88    \\
27 Aug 1998 & 0.17$\pm$0.13 & -4.83$\pm$3.75    \\
28 Aug 1998 & 0.18$\pm$0.10 & -5.03$\pm$2.83    \\
1 Aug 1999  & 0.01$\pm$0.29 & -3.66$\pm$8.19    \\
2 Aug 1999  & 0.06$\pm$0.11 & -1.21$\pm$3.43    \\
3 Aug 1999  & -0.05$\pm$0.18&  3.00$\pm$5.32    \\
5 Aug 1999  & 0.14$\pm$0.24 & -2.80$\pm$7.41    \\
\enddata
\end{deluxetable}

\end{document}